\documentclass[10pt,twocolumn,letterpaper]{article}

\usepackage{cvpr}              
\usepackage[accsupp]{axessibility}

\usepackage{graphicx}
\usepackage{latexsym}
\usepackage[pagebackref,breaklinks,colorlinks,citecolor=cvprblue]{hyperref}
\usepackage{prettyref}
\usepackage{amsmath}
\usepackage{amsfonts}
\usepackage{amssymb}
\usepackage{amsthm}
\usepackage{multirow}
\usepackage{booktabs}
\usepackage{algorithmic}
\usepackage{algorithm}
\usepackage{array}
\usepackage{textcomp}
\usepackage{stfloats}
\usepackage{verbatim}
\usepackage{graphicx}
\usepackage{mathtools}

\usepackage{longtable}

\newcommand{\pref}[1]{\prettyref{#1}}

\newrefformat{sec}{Sec.~\ref{#1}}
\newrefformat{apx}{Appendix~\ref{#1}}
\newrefformat{lem}{Lemma~\ref{#1}}
\newrefformat{fig}{Fig.~\ref{#1}}
\newrefformat{alg}{Algorithm~\ref{#1}}
\newrefformat{table}{Table~\ref{#1}}

\def\Dt{\Delta{t}}

\def\R{\mathbb{R}}
\def\loss{\mathcal{L}}

\def\etal{et al.}

\def\p{\boldsymbol{p}}

\def\v{\boldsymbol{v}}

\def\a{\boldsymbol{a}}
\def\b{\boldsymbol{b}}
\def\n{\boldsymbol{n}}

\def\bbeta{\boldsymbol{\beta}}

\def\cov{\boldsymbol{\Sigma}}
\def\shape{\boldsymbol{\beta}}
\def\pose{\boldsymbol{\theta}}

\newcommand{\norm}[1]{\left\lVert#1\right\rVert}

\definecolor{cvprblue}{rgb}{0.21,0.49,0.74}

\def\paperID{14944}
\def\confName{CVPR}
\def\confYear{2025}

\title{
UMotion: Uncertainty-driven Human Motion Estimation from \\Inertial and Ultra-wideband Units 
}

\author{Huakun Liu\qquad Hiroki Ota\qquad Xin Wei\qquad Yutaro Hirao\\
Monica Perusqu\'ia-Hern\'andez\qquad Hideaki Uchiyama\qquad Kiyoshi Kiyokawa\\
Nara Institute of Science and Technology, Japan\\
{\tt\small {\{liu.huakun.li0, ota.hiroki.oc6, wei.xin.wy0, yutaro.hirao\}}@is.naist.jp}\\
{\tt\small {\{m.perusquia, hideaki.uchiyama, kiyo\}}@is.naist.jp}
}

\begin{document}
\maketitle
\begin{abstract}

Sparse wearable inertial measurement units (IMUs) have gained popularity for estimating 3D human motion.
However, challenges such as pose ambiguity, data drift, and limited adaptability to diverse bodies persist.
To address these issues, we propose UMotion, an uncertainty-driven, online fusing-all state estimation framework for 3D human shape and pose estimation, supported by six integrated, body-worn ultra-wideband (UWB) distance sensors with IMUs.
UWB sensors measure inter-node distances to infer spatial relationships, aiding in resolving pose ambiguities and body shape variations when combined with anthropometric data.
Unfortunately, IMUs are prone to drift, and UWB sensors are affected by body occlusions.
Consequently, we develop a tightly coupled Unscented Kalman Filter (UKF) framework that fuses uncertainties from sensor data and estimated human motion based on individual body shape.
The UKF iteratively refines IMU and UWB measurements by aligning them with uncertain human motion constraints in real-time, producing optimal estimates for each.
Experiments on both synthetic and real-world datasets demonstrate the effectiveness of UMotion in stabilizing sensor data and the improvement over state of the art in pose accuracy.
Code is available at: \url{https://github.com/kk9six/umotion}.
\end{abstract}

\section{Introduction}
\label{sec:intro}
\begin{figure}
    \centering
    \includegraphics[width=.8\linewidth]{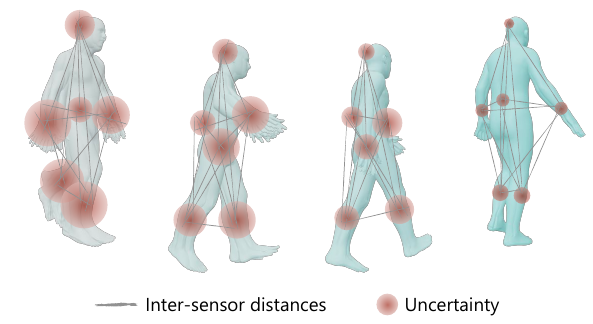}
    \caption{UMotion integrates IMU-UWB data inputs and pose outputs uniformly under uncertainty, constrained by individual body structure.
    The online state estimation framework iteratively refines sensor data confidence and stabilizes pose estimation, reducing ambiguities and improving robustness.}
    \label{fig:teaser}
\end{figure}
Estimating 3D human motion from wearable sensors has become increasingly popular due to their portability, accessibility, and versatility.
Wearable sensors, such as inertial measurement units (IMUs), enable continuous monitoring of body motion measurements across unrestricted spaces.
These advances shift motion capture from controlled laboratory environments to everyday settings~\cite{von2017sparse, mollyn2023imuposer, zuo2024loose, lee2024mocap}.
This transition benefits fields such as healthcare, sports performance, ergonomics, and emerging areas in human-computer interaction~\cite{jiang2022avatarposer, van2023diffusion, lee2023questenvsim, dai2024hmd}.

One of the widely chosen wearable sensors for 3D human motion estimation is IMU.
Commercial systems, such as Xsens~\cite{schepers2018xsens}, utilize 17 or more IMUs for comprehensive pose coverage.
While highly accurate, these densely placed IMUs are inconvenient and intrusive.
Recent studies have reduced the required number of IMUs to just six---placed on the forearms, lower legs, pelvis, and head---while still achieving promising performance through data-driven methods~\cite{von2017sparse, huang2018deep,yi2021transpose,yi2022physical,jiang2022transformer,van2023diffusion,zhang2024dynamic,yi2024physical,armani2024ultra}.
With fewer IMUs, pose estimation becomes under-constrained and prone to ambiguity.
Recent work has attempted to disambiguate poses by incorporating temporal consistency, physics-based constraints, or additional, easy-to-integrate sensors~\cite{yi2022physical,jiang2022transformer, armani2024ultra}.
However, these methods still face significant challenges, including noisy sensor data, body shape variations, and ambiguities arising from under-constrained setups with sparse sensors.

In this work, we propose UMotion, an uncertainty-driven human motion estimation framework that combines online state estimation with an integrated system of six inertial and ultra-wideband (UWB) distance sensors.
Prior work often infers poses from sensor data without considering the reverse influence.
UMotion uniformly treats input and output under uncertainty and body-specific constraints, blending data to optimize estimates (see ~\pref{fig:teaser}).
Our approach maintains inter-sensor distances as a core system state, capturing spatial relationships between sensor nodes.
These distances, combined with basic anthropometric measurements (height and weight) and IMU data, serve as inputs for our learning-based shape and pose estimators.
Since sensors are inherently noisy, we propose a tightly coupled online state estimation system that associates IMU and UWB sensor measurements and pose estimates using an Unscented Kalman Filter (UKF) and an uncertainty propagation method, iteratively correcting errors and stabilizing pose estimation in real-time.
Our main contributions are:
\begin{itemize}
    \item An ensemble learning-based human shape estimation approach using distance constraints and anthropometric measurements from six integrated IMU-UWB sensors.
    \item A learning-based method for human pose estimation from inertial and distance constraints, incorporating mesh distribution estimation.
    \item A filtering-based state estimation system that couples sensor measurements, pose estimates, and body variations in real-time to improve sensor stability and pose accuracy.
\end{itemize}

\section{Related Work}
\label{sec:related}
\paragraph{Pose Estimation from Sparse Inertial Sensors}
In contrast to the predominant use of vision-based methods, wearable sensors, specifically IMUs, offer greater freedom and flexibility.
Von Marcard~\etal~\cite{von2017sparse} present SIP that makes inertial pose estimation practical by using only six IMUs attached to the body, combined with an offline iterative SMPL body model~\cite{loper2023smpl} pose optimization.
Huang~\etal~\cite{huang2018deep} propose a real-time pose estimation method, DIP, which uses a bidirectional recurrent neural network (biRNN)~\cite{schuster1997bidirectional} to learn the mapping from a sequence of six IMU measurements to SMPL body poses.
To obtain sufficient training data, they synthesize IMU measurements from the AMASS dataset~\cite{mahmood2019amass}, further advancing the development of learning-based methods for sparse sensor motion capture.

Following previous studies, TransPose~\cite{yi2021transpose} refines pose estimation by decomposing the end-to-end framework into a multi-stage process with intermediate joint position estimation.
To disambiguate the poses, PIP~\cite{yi2022physical} proposes a physics-based motion optimization, while TIP~\cite{jiang2022transformer} incorporates a conditional Transformer model for plausible terrain generation. 
Training data is crucial for learning-based methods.
Unlike prior studies relying on synthetic IMU data, DynalIP~\cite{zhang2024dynamic} adapts real IMU data from diverse human skeleton formats to the target SMPL model and shows superior performance when training with real sensor data.
Similarly focusing on the data, PNP~\cite{yi2024physical} recently addresses limitations in existing IMU synthesis by incorporating non-inertial effects and fictitious forces, enhancing estimation robustness through a physics-informed neural network and realistic IMU synthesis.

In contrast to methods focusing solely on pose estimation, our approach treats human shape and pose as equally important.
By integrating shape information, we establish a tight connection between sensor data and estimated motions, forming a positive feedback loop that enhances the entire process.
Additionally, most previous studies fail to fully utilize IMU accelerations due to high noise and drift issues.
Within our framework, accelerations serve as control inputs for state estimation, undergoing continuous correction and acting as a crucial component in tracking spatial relationships among sensors.

\paragraph{Pose Estimation from Hybrid Sensors}
In addition to IMUs, various wearable and hybrid sensor systems have been explored to overcome inherent IMU limitations such as restricted positional awareness~\cite{pons2010multisensor, liu2011realtime, kaufmann2021pose, jiang2023egoposer, liang2023hybridcap, pan2023fusing, ponton2023sparseposer, yi2023egolocate, ren2023lidar, lee2024mocap}.
The closest works to ours are SmartPoser~\cite{devrio2023smartposer} and Ultra Inertial Poser (UIP)~\cite{armani2024ultra}, both of which integrate UWB and IMU sensors for pose estimation.
UWB sensors complement IMUs with additional distance information between sensors while preserving the portability and flexibility of tracking devices.
SmartPoser~\cite{devrio2023smartposer} focuses on wearable arm pose estimation, combining UWB measurements with IMU data using an off-the-shelf smartwatch and smartphone.
UIP~\cite{armani2024ultra} integrates six UWB sensors with IMUs for full-body pose estimation.
Inter-sensor distances, estimated using an EKF that fuses IMU and UWB data, help reduce global translation drift and minimize position jitter compared to inertial-only tracking.
However, body occlusion frequently affects certain node pairs, e.g., head-knee, causing the EKF to fail in estimating distances due to rapid IMU drift and unreliable UWB measurements, resulting in unstable distance data.
Our approach addresses this limitation by using output pose uncertainties and body-specific constraints as observations to refine input sensor measurements, improving the stability of both IMU and UWB data.

\section{Method}
\label{sec:method}
\subsection{Preliminaries}
\noindent\textbf{SMPL body model}\quad
We use the SMPL model~\cite{loper2023smpl} to represent human motion.
SMPL decomposes the human body into pose parameters, $\pose \in \mathbb{R}^{23 \times 3 + 3}$, which define relative rotations of 23 joints and the global root joint orientation in axis-angle form, and shape parameters, $\shape \in \mathbb{R}^{10}$, capturing body shape variations across individuals.
The model is defined by a linear blend-skinning (LBS) function:
\begin{equation}\label{eq:smpl}
    M(\bbeta, \pose) = W\left(\mathbf{\bar{T}} + B_s(\bbeta) + B_p(\pose), J(\bbeta), \pose, \mathcal{W}\right),
\end{equation}
where $B_s(\bbeta)$ and $B_p(\pose)$ are shape and pose blend shapes that deform the template mesh $\mathbf{\bar{T}}$ in the zero pose.
The mesh is then reposed according to specified poses $\pose$, combining with joint positions $J(\bbeta)$ and blend-skinning weights $\mathcal{W}$.
\noindent\textbf{Inertial Measurement Unit (IMU)}\quad
A nine-axis MEMS IMU comprises an accelerometer, gyroscope, and magnetometer, measuring accelerations in the sensor local frame $F^S$, angular velocities in $F^S$, and magnetic field strengths relative to the Earth’s magnetic field, respectively.
A commonly used error model for measured accelerations of one IMU, $\a \in \R^{3}$, is given by~\cite{ru2022mems}:
\begin{equation}\label{eq:imu_model}
\a = \a_{\text{true}} + \b_{a} + \n_{a},
\end{equation}
where $\a_{\text{true}}$ is the true acceleration, $\b_{a}$ is the bias, modeled as a random walk process with noise $\boldsymbol{\eta}$ following a zero-mean Gaussian distribution with standard deviation $\boldsymbol{\Sigma}_b$, and $\n_{a}$ is high-frequency white noise.
In our work, after a frame calibration process, accelerations are represented as $\a^{M}$ in the SMPL body-centric frame $F^M$, and orientations are represented as $\boldsymbol{R}^{MB}$, indicating the rotation from bone frame $F^{B}$ to $F^M$.
For simplicity, we denote $\a$ as $\a^{M}$ and $\boldsymbol{R}$ as $\boldsymbol{R}^{MB}$ in the following sections.
Details on the calibration process are available in supplementary materials.

\noindent\textbf{Ultra-wideband (UWB)}\quad
UWB uses the time-of-flight technique to measure distances between two devices, with one acting as the transmitter and the other as the receiver.
Unlike other RF techniques, such as WiFi and Bluetooth, UWB operates over a wide frequency range and transmits short pulses, achieving centimeter-level distance measurements with minimal interference~\cite{oppermann2004uwb}.
Consequently, it is widely used in high-precision indoor localization systems with multiple fixed anchors and movable tags~\cite{zafari2019survey}.
In our work, all six integrated UWB sensors are movable and alternately function as both transmitters and receivers, capturing inter-distances between each sensor pair.

\subsection{Framework Design}
\begin{figure*}
    \centering
    \includegraphics[width=.9\textwidth]{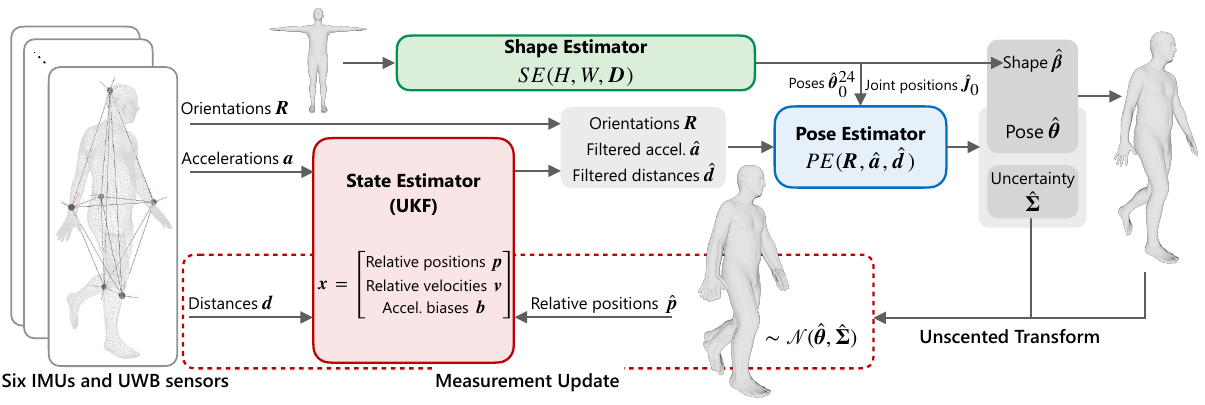}
    \caption{Overview of UMotion, consisting of three main modules: the shape estimator, pose estimator, and state estimator.
    The shape estimator takes anthropometric measurements and inter-distances in a T-pose as input, outputting shape parameters that reconstruct a realistic body and impose strong constraints on the system.
    The pose estimator receives filtered IMU data and inter-distances from the state estimator to predict poses, which are fed back to refine state estimates.
    The entire system integrates IMU, UWB, and estimated poses within the context of individual body structure to continuously update and improve motion estimation.
    }
    \label{fig:overview}
\end{figure*}
As shown in~\pref{fig:overview}, our method uses six IMU-UWB sensors and consists of three main modules: a shape estimator, a pose estimator, and a state estimator.
The shape estimator takes anthropometric measurements and inter-distances in a T-pose as input, outputting shape parameters $\hat\bbeta$ of the SMPL model.
The pose estimator uses filtered frame-aligned, root-normalized IMU measurements and inter-distances as input to estimate pose parameters, $\hat\pose$, and corresponding uncertainties, $\hat{\boldsymbol{\Sigma}}$.
Given our sensor placement, we do not observe the movement of the hands and feet.
Therefore, similar to previous studies, we estimate poses for only 16 joints, including the root joint, while excluding the hand and foot joints.
For simplicity, we denote this reduced set of pose parameters as $\pose \in \mathbb{R}^{16 \times 3}$, while $\pose^{24}$ refers to the full set of pose parameters for 24 joints.
Notations with a hat symbol, e.g., $\hat{\pose}$, indicate the corresponding estimated values.
Additionally, we do not estimate global translation, as inferring it from only body-worn IMU and UWB sensors without external references would lead to unbounded error accumulation.
The state estimator integrates accelerations, UWB measurements, and the estimated poses with uncertainties to track the sensor relative positions, velocities, and acceleration biases of each node at each time step.
These refined estimates are then used to refine the inputs to the pose estimator.

The estimation framework is driven by inherent uncertainties present in the system.
Specifically, noisy IMU data combined with tracked inter-sensor distances serve as the primary inputs for pose estimation. 
The generated poses, along with their associated uncertainties that reflect sensor noise, are further constrained by the human model, which in turn refines sensor measurements and distance estimates.
This feedback loop propagates uncertainties through the system, continuously updating the belief in system states based on the reliability of each data source—IMU, UWB, and pose estimator with body constraints, ultimately producing robust and accurate estimation.

\subsection{Shape and Pose Estimator}
\subsubsection{Shape Estimator}
\begin{figure}[t]
    \centering
    \includegraphics[width=0.4\textwidth]{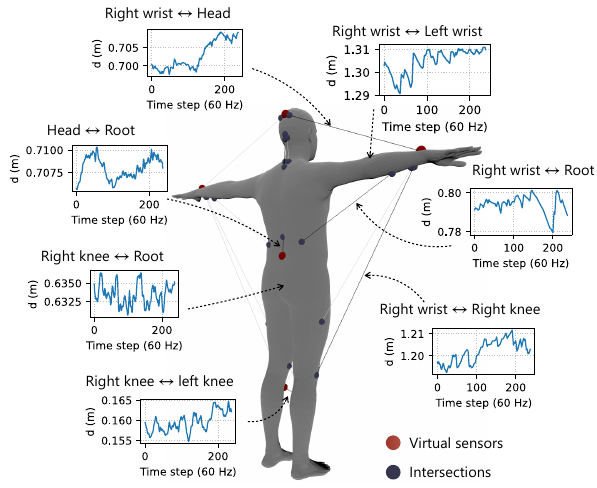}
    \caption{Visualization of selected inter-sensor distances used in the shape estimator.
    We place virtual sensors on the body mesh at sensor mounting points and conduct line-of-sight simulation experiments.
    The plots display the temporal changes in UWB measurements between various sensor pairs over time.}
    \label{fig:t-pose}
\end{figure}
Spatial inter-distances provide data that approximates the rough skeletal structure of an individual.
Previous studies, such as Virtual Caliper~\cite{pujades2019virtual} and SHAPY~\cite{choutas2022accurate}, have demonstrated a linear relationship between body shape and body measurements.
Building on these findings, we adopt an ensemble learning-based shape estimator that uses experimentally selected anthropometric measurements, defined as:
\begin{equation}
   \hat\shape = SE(H, W, \boldsymbol{D}).
\end{equation}
Basic body measurements, height $H \in \mathbb{R}^1$  and weight $W \in \mathbb{R}^1$, are easily accessible and provide foundational information about human body proportions.
Experimentally selected inter-distances between sensors, $\boldsymbol{D} \in \R^7$,  capture relative limb lengths and body structural details that height and weight alone cannot provide.

As shown in~\pref{fig:t-pose}, we place virtual sensors on the body mesh and conduct line-of-sight (LOS) simulation experiments with real-world tests to identify and select 7 repeatable distances out of 15 possible options.
While the selection may vary based on device characteristics, the chosen distances offer a reliable foundation for adaptation.
For model training, we use AutoGluon~\cite{erickson2020autogluon}, an AutoML framework that automatically trains and ensembles 11 basic machine learning models, which is more stable than a single linear regression method in our validation.
The estimated shape parameters $\hat\shape$ are then used to reconstruct realistic human bodies and to propagate estimated poses to spatial constraints in the state estimator.

\subsubsection{Pose Estimator}
Our pose estimator is based on unidirectional multi-layer long short-term memory (LSTM)~\cite{hochreiter1997long} recurrent neural network (RNN).
The unidirectional LSTM preserves only past information, achieving real-time estimation without requiring future data.
Given six IMU-UWB sensors, we use filtered accelerations, $\hat{\a} \in \R^{18}$, and orientations in 9D rotation matrix form, $\boldsymbol{R} \in \R^{54}$, along with 15 inter-distances $\hat{\boldsymbol{d}} \in \R^{15}$, as inputs.
The estimator then outputs the poses $\hat{\pose}_{6D} \in \R^{96}$ for the target joints, represented in a 6D rotation matrix form~\cite{zhou2019continuity}.
Additionally, in the final layer, the network outputs the logarithm of the estimation uncertainty, $\hat{\cov} \in \mathbb{R}^{96}$, capturing prediction confidence.
Consequently, the pose estimator is defined as:
\begin{equation}
    \hat{\pose}_{6D}, \hat{\cov} = PE(\hat{\a}, \boldsymbol{R}, \hat{\boldsymbol{d}}).
\end{equation}
To handle ambiguity in the initial frame, we adopt a learning-based initialization strategy inspired by PIP~\cite{yi2022physical}.
A fully-connected convolutional neural network regresses the initial hidden state of the RNN based on the initial global poses, $\hat{\pose}^{24}_0$, and joint positions, $\hat{\boldsymbol{J}}_0 \in \mathbb{R}^{72}$.
This approach provides the pose estimator with a well-informed initialization, adapting to different bodies.

We use two loss functions to train the pose estimator: Mean Squared Error (MSE) and Gaussian Negative Log Likelihood (GNLL) loss.
MSE is defined as:
\begin{equation}
    \loss_\text{MSE}(\pose_{6D}, \hat{\pose}_{6D}) = \frac{1}{n}\sum^n_{i=1}\norm{\pose_{6D} - \hat{\pose}_{6D}}^2_2,
\end{equation}
where $\pose_{6D}$ denotes the ground truth poses and $n$ is the total number of frames.
The GNLL, $\loss_\text{GNLL}(\pose_{6D}, \hat{\pose}_{6D}, \hat{\cov}) = $
\begin{equation}
    \frac{1}{2}\left(\log\left(\max\left(\hat{\cov}^2, \boldsymbol{\varepsilon}_{\min}\right)\right) + \frac{(\hat{\pose}_{6D} - \pose_{6D})^2}{\max\left(\hat{\cov}^2, \boldsymbol{\varepsilon}_{\min}\right)}\right),
\end{equation}
where $\boldsymbol{\varepsilon}_{\min}$, set to $10^{-6}$ in our implementation, is used for numerical stability.
First, $\loss_\text{MSE}$ helps the model converge quickly during early training.
Next, $\loss_\text{GNLL}$ is applied to optimize for pose and uncertainty estimation.

\subsection{State Estimator}
We use UKF~\cite{wan2000unscented} for state estimation to refine the pose estimator inputs, integrating UWB measurements and pose estimator outputs with a statistically derived IMU model.
The UWB measurements provide direct inter-distance data, while they are affected by body occlusion issues.
Meanwhile, pose estimator outputs offer constraints and inform distance measurements; however, they suffer from occasional inaccuracies due to challenging poses or sensor noise.
The UKF combines these sources, balancing the strengths and limitations of each to make the optimal estimates.
\subsubsection{State Definition}
We define the state vector $\boldsymbol{x} \in \mathbb{R}^{15\times 3 + 15\times 3 + 6\times 3}$ as:
\begin{equation}
\boldsymbol{x} = 
\begin{bmatrix}
\p^{12} & \dots & \p^{56} & \v^{12}& \dots & \v^{56} & \b^1 & \dots & \b^6
\end{bmatrix}^\mathsf{T},
\end{equation}
where $\p^{xy}$ represents the relative position between nodes $x$ and $y$, with $x,y \in \{1, 2, 3, 4, 5, 6\}$ and $x < y$. The term $\v^{xy}$ denotes the relative velocity between nodes $x$ and $y$, while $\b^1$ through $\b^6$ represent residual acceleration errors after transformation from $F^{S}$ to $F^{M}$. These biases primarily result from orientation errors and raw accelerometer biases $\b_{a}$.
All quantities are expressed in the SMPL body-centric coordinate frame $F^M$.
For initialization, we set $\p^{xy}$ based on a static T-pose defined by the shape parameters $\shape$ and the SMPL model~\pref{eq:smpl}.
The relative velocities \(\v^{xy}\) and biases are initialized to zero.

\subsubsection{State Propagation}
The control input $\boldsymbol{u}$ consists of the accelerations $\a$ of six nodes, defined as:
\begin{equation}
    \boldsymbol{u} = \begin{bmatrix}\a^{1}  & \dots & \a^{6} \end{bmatrix}^\mathsf{T}, \boldsymbol{u} \in \mathbb{R}^{18}.
\end{equation}
During the prediction step, the current state $\boldsymbol{x}$ is propagated using the control input $\boldsymbol{u}$.
The propagation model is derived from the strapdown inertial kinematic model with an acceleration error model~\pref{eq:imu_model}, defined as follows:
\begin{align}
\v_{k}^{xy} &= \v_{k-1}^{xy} + (\a_{k-1}^{y} - \a_{k-1}^{x})\Dt - (\b_{k-1}^{y} - \b_{k-1}^{x})\Dt, \label{eq:v_xy}\\
\p_{k}^{xy} &= \p_{k-1}^{xy} + \v_{k-1}^{xy}\Dt + \frac{1}{2}(\a_{k-1}^{y} - \a_{k-1}^{x}) \Dt^2 \nonumber \\ &\qquad+\frac{1}{2}(\b_{k-1}^{x} - \b_{k-1}^{y})\Dt^2, \label{eq:p_xy}\\
\b^x_{k} &= \b^x_{k-1} + \boldsymbol{\eta}^x, \boldsymbol{\eta}^x\sim\mathcal{N}(0, \cov_b^2) \label{eq:b_x},
\end{align}
where $k-1$ and $k$ denote consecutive time steps, and $\Dt$ represents the interval between them.
Together, equations~\pref{eq:v_xy}, \pref{eq:p_xy}, and \pref{eq:b_x} form a state propagation model with added white noise, represented as:
\begin{equation}
\bar{\boldsymbol{x}}_{k} = f(\boldsymbol{x}_{k-1}, \boldsymbol{u}_{k}) + \mathbf{Q},
\end{equation}
where $\mathbf{Q}$ is the process noise covariance matrix, derived from the characteristics of the IMUs used in the system.

\subsubsection{Measurement Update}
The state is updated within measurement spaces using observed data $\boldsymbol{z}$.
To accomplish this, we define a measurement model $h(\boldsymbol{x})$ that maps the current state $\boldsymbol{x}$ to measurement spaces as:
\begin{equation}
    h(\boldsymbol{x}) = \begin{bmatrix}
        \norm{\p^{xy}}_2 & \norm{\v^{xy}}_2 & \p^{xy} & \v^{xy}
    \end{bmatrix}^\mathsf{T}.
\end{equation}
The UWB sensor measures the distance, $d^{xy}$, between node $x$ and $y$, which corresponds to the distance derived from relative positions in state, $\norm{\p^{xy}}_2$.
Additionally, from consecutive distance measurements, we derive the relative velocity norm as $({d}^{xy}_{k} - {d}^{xy}_{k-1}) / \Dt$, corresponding to $\norm\v^{xy}_2$.
The covariance matrix for distance measurements at time step $k$, denoted as $\mathbf{R}_{1, k}$, is dynamically set based on line-of-sight conditions for each node pair, informed by $\hat{\pose}_{k}$ and UWB sensor characteristics.
The covariance matrix for relative velocity measurements, $\mathbf{R}_{2, k}$, is computed as:
\begin{equation}
    \mathbf{R}_{2,k} = \frac{(\mathbf{R}_{1, k} + \mathbf{R}_{1, k-1})}{\Dt^2}.
\end{equation}

The pose estimator outputs the joint rotations, $\hat\pose$, along with their corresponding standard deviations, $\hat\cov$.
We employ the unscented transform~\cite{julier1995new} to transform the pose distribution to the relative position distribution.
Given $\boldsymbol{\Theta} \in \R^{96} \sim \mathcal{N}(\hat{\pose}, \hat{\cov})$, we use Van der Merwe’s scaled sigma point algorithm~\cite{van2004sigma} to generate $n$ sigma points, $\boldsymbol{\mathcal{X}} \in \mathbb{R}^{n \times 96}$, along with weights $\boldsymbol{W}^m \in \mathbb{R}^{n}$ for the mean and $\boldsymbol{W}^c \in \mathbb{R}^{n}$ for the covariance.
Each sigma point in $\boldsymbol{\mathcal{X}}$ is transformed through the SMPL model~\pref{eq:smpl}, yielding a set of sensor-relative positions denoted as:
\begin{equation}
\mathcal{Y} = \{\p^{xy}_i \in \R^{n\times 15 \times 3} \mid i = 1, 2, \dots, n \}.
\end{equation}
We then calculate the mean and covariance of the transformed relative positions, capturing the distribution of relative positions based on pose uncertainty:
\begin{align}
    \hat{\p}^{xy} &= \sum_{i=1}^{n} \boldsymbol{W}^m_i\mathcal{Y}_i, \\
   \hat{\cov}_{\hat\p}^2 &= \sum_{i=1}^{n} \boldsymbol{W}^c_i(\mathcal{Y}_i - \hat{\p}^{xy})(\mathcal{Y}_i-\hat{\p}^{xy})^\mathsf{T}.
\end{align}
The transformed relative positions, $\hat{\p}^{xy}$, along with the measurement covariance matrix $\mathbf{R}_3 = \hat{\cov}_{\hat{\p}}^2$, are used to update $\p^{xy}$ in the state $\boldsymbol{x}$.
Similarly, we compute the relative velocity between consecutive time steps as $(\hat{\p}^{xy}_k - \hat{\p}^{xy}_{k-1}) / \Dt$, with its covariance matrix defined as $\mathbf{R}_{4, k} = (\mathbf{R}_{3, k} + \mathbf{R}_{3, k-1})/\Dt^2$, to update $\v^{xy}$. 
The complete measurement vector $\boldsymbol{z}$ used for updating the state is given by:
\begin{equation}
    \boldsymbol{z} = \begin{bmatrix}
        {d}^{xy} &  \frac{({d}^{xy}_{k} - {d}^{xy}_{k-1})}{\Dt} & \hat{\boldsymbol{p}}^{xy} & \frac{(\hat{\p}^{xy}_k - \hat{\p}^{xy}_{k-1})}{\Dt}
    \end{bmatrix}^\mathsf{T}.
\end{equation}

\section{Experiments}
\subsection{Datasets}
We use the AMASS~\cite{mahmood2019amass}, TotalCapture~\cite{trumble2017total}, DIP-IMU~\cite{huang2018deep}, and UIP~\cite{armani2024ultra} datasets for training and evaluation.
We synthesize inertial and distance data from the AMASS dataset, following the process in previous studies~\cite{huang2018deep, yi2021transpose, armani2024ultra}.
We select specific vertices on the body mesh to represent sensor mounting positions.
Accelerations are computed as finite differences of these vertex positions over time, while orientations are obtained as the global orientations of the corresponding joints.
Distance measurements are synthesized by calculating Euclidean distances between selected sensor positions, adjusted for individual body shape.
Training the shape estimator requires anthropometric measurements and inter-distances with corresponding shape parameters from a large number of individuals, which is challenging to collect.
Therefore, we synthesize these measurements—height $H$, weight $W$, and inter-distances $\boldsymbol{D}$—from the 3D body mesh in a T-pose, using 479 unique body shapes (273 male and 206 female) available in the AMASS dataset.
We estimate $H$ and $W$ following the methods in SHAPY~\cite{choutas2022accurate}, and we sample $\boldsymbol{D}$ from the synthesized distances.

The training data for the pose estimator comprises synthesized data from the AMASS dataset and the training set of DIP-IMU.
Unlike previous studies~\cite{yi2021transpose, yi2022physical, jiang2022transformer}, we exclude the TotalCapture within AMASS from the training data to prevent overlap, ensuring that synthesized distances used in testing are unseen during training.
The test data includes TotalCapture, the test split of DIP-IMU, and UIP.

\subsection{Method Implementation}
Our shape estimator was implemented using the official multi-label predictor in AutoGluon 1.1 with a medium-quality preset.
Our pose estimator was implemented based on PyTorch 2.3.
The training process used an Adam optimizer with a learning rate of 0.0001.
The pose estimator model was trained for 350 epochs with a batch size of 512 on a single NVIDIA GeForce RTX 3090 GPU, completing in approximately 35 minutes.
For initial 20 epochs, we adopted $\loss_{MSE}$ to optimize the model, then switched to $\loss_{GNLL}$ for the remaining epochs.
For the UKF, we experimentally set parameters $\alpha_{UKF} = 0.2$, $\beta_{UKF} = 1.0$, and $\kappa_{UKF} = -105$ to control the distribution and weighting of sample points.
In the unscented transform applied to pose estimates, we used $\alpha_{NN} = 0.09$, $\beta_{NN} = 1.0$, and $\kappa_{NN} = -93$.
We applied a factor of 10 to $\mathbf{R}_3$ to compensate for overconfident predictions.
The method achieves 60~Hz without online LOS inference and 30~Hz with it.
\subsection{Quantitative Evaluation}
We evaluate our method against two categories of methods: (1) IMU-only methods, including DIP~\cite{huang2018deep}, TransPose~\cite{yi2021transpose}, TIP~\cite{jiang2022transformer}, PIP~\cite{yi2022physical}, and PNP~\cite{yi2024physical}, and (2) distance augmented methods, including TIP-D, PIP-D, and UIP~\cite{armani2024ultra}.
TIP-D and PIP-D are modified versions of TIP and PIP with distance-augmented input data, and we use their results as reported by UIP.
We follow the evaluation protocol in UIP by adding ideal synthetic inter-sensor distances without noise to TotalCapture and DIP-IMU to assess the impact of distance measurements independent of noise.
For evaluations on the UIP dataset, which includes both IMU and UWB data, we perform RANSAC regression to calibrate UWB measurements, as described in UIP~\cite{armani2024ultra}, and then use calibrated measurements as input for the state estimator.
We do not apply outlier filtering and use default state estimator parameters due to the absence of ground-truth data and limited information on sensor-specific characteristics in the hardware setup.

\noindent\textbf{Metrics}\quad
We use the following metrics for quantitative evaluation: 1) \textit{SIP Error (in degrees)}: the mean global angular error of the upper arms and upper legs, focusing on joints not observed by the body-worn sensors; 2) \textit{Angular Error (in degrees)}: the mean global angular error across all body joints; 3) \textit{Positional Error (in centimeters)}: the mean global joint position error across all joints; and 4) \textit{Mesh Error (in centimeters)}: the mean vertex position error between the reconstructed mesh and the ground-truth mesh given the mean body shape.

\noindent\textbf{Comparison with IMU-only Methods}\quad
\begin{table*}[t]
    \small
    \centering
    \begin{tabular}{ccccccccc}
        \toprule
        \multirow{2}{*}{Method} & SIP Error & Ang Error & Pos Error & Mesh Error &  SIP Error & Ang Error & Pos Error & Mesh Error \\
        & (deg) & (deg) & (cm) & (cm) & (deg) & (deg) & (cm) & (cm) \\
        \midrule
        \multicolumn{1}{c}{} &  \multicolumn{4}{c}{TotalCapture} & \multicolumn{4}{c}{DIP-IMU} \\
        \midrule
        DIP~\cite{huang2018deep} & 18.62 & 17.22 & 9.42 & 11.22  & 17.35 & 15.36 & 7.59 & 9.05 \\
        TransPose~\cite{yi2021transpose} & 16.58 & 12.89 & 6.55 & 7.42  & 17.06 & 8.86 & 6.03 & 7.17  \\
        TIP~\cite{jiang2022transformer} & 13.22 & 12.30 & 5.81 & 6.80 & 16.90 & 9.07 & 5.63 & 6.62  \\
        PIP~\cite{yi2022physical} & 12.93 & 12.04 & 5.61 & 6.51 & 15.33 & 8.78 & 5.12 & 6.02  \\
        PNP~\cite{yi2024physical} & 10.89 & 10.45 & 4.74 & 5.45 & \textbf{13.71} & 8.75 & 4.97 & 5.77  \\
        \textbf{UMotion} & \textbf{10.76} & \textbf{7.06} & \textbf{4.46} & \textbf{4.94} & 14.19 & \textbf{6.35} & \textbf{3.38} & \textbf{3.93} \\
        \bottomrule
    \end{tabular}
    \caption{Comparison with state of the art IMU-only methods on TotalCapture~\cite{trumble2017total} and DIP-IMU~\cite{huang2018deep}.}
    \label{table:com_inertial}
\end{table*}
\pref{table:com_inertial} presents a comparison of our method against IMU-only methods on the TotalCapture and DIP-IMU datasets.
On the TotalCapture dataset, our approach consistently outperforms previous methods across all metrics, improving over the SOTA PNP by 32.4\% in angular error and 9.4\% in mesh error.
A similar trend is also observed in the results on DIP-IMU dataset.
This demonstrates that inter-sensor distances serve as strong constraints for refining estimated poses.

\noindent\textbf{Comparison with Distance-augmented Methods}
\begin{table*}[t]
    \small
    \centering
    \begin{tabular}{ccccccc}
        \toprule
        \multirow{2}{*}{Method} & SIP Error &  Pos Error & SIP Error & Pos Error & SIP Error & Pos Error \\
               & (deg) & (cm) & (deg) & (cm) & (deg) & (cm) \\
        \midrule
        \multicolumn{1}{c}{} & 
        \multicolumn{2}{c}{TotalCapture} &
        \multicolumn{2}{c}{DIP-IMU} & 
        \multicolumn{2}{c}{UIP}\\
        \midrule
        PIP~\cite{yi2022physical} & 15.93 & 7.05 & 15.98 & 6.21 & 30.47 & 13.62 \\
        \midrule
        TIP-D & 12.18 & 5.51 & 15.91 & 5.26 & 30.34 & 13.96 \\
        PIP-D & 13.16 & 6.31 & 13.79 & 5.36 & 30.33 & 13.27 \\
        UIP~\cite{armani2024ultra}   & 11.32 & 5.49 & \textbf{13.21} & 5.05 & \textbf{24.12} & 10.65 \\
        \textbf{UMotion}  & \textbf{10.76} & \textbf{4.46} & 14.19 & \textbf{3.38} & 25.69 & \textbf{10.33} \\
        \bottomrule
    \end{tabular}
    \caption{Comparison with distance-augmented methods on TotapCapture~\cite{trumble2017total}, DIP-IMU~\cite{huang2018deep}, and UIP~\cite{armani2024ultra} datasets.}
    \label{table:com_inertial_distance}
\end{table*}
\pref{table:com_inertial_distance} presents a fair comparison of our method with distance-augmented methods, as all approaches use the same input data.
Additionally, all methods are trained on the AMASS dataset with the TotalCapture subset excluded, whereas the IMU-only methods in~\pref{table:com_inertial} include TotalCapture in their training data.
Our method outperforms UIP, achieving a reduction in positional error of 21.5\% on the TotalCapture dataset and 35.0\% on the DIP-IMU dataset.
On the UIP dataset, where sensor data quality is low and motions are more ambiguous than in the other two test datasets, our method achieves the lowest positional error, while UIP achieves the lowest SIP error.
We achieve comparable results despite using less-filtered distance measurements, a simpler architecture, and non-optimal default parameters for the state estimator.
This demonstrates the robustness of our approach in handling real-world sensor noise and the effectiveness of our online state estimation framework.

\subsection{Module Evaluations}
\noindent\textbf{Shape Estimator}\quad
\begin{table}[t]
    \small
    \centering
    \begin{tabular}{cccc}
        \toprule
        Metric & Mean shape & Predicted &  GT \\
        \midrule
        Pos Error (cm) & 5.15  & 4.34 & 4.31 \\
        Mesh Error (cm) & 5.63 & 4.81 & 4.78\\
        $H$ Error (cm) & 5.75 & 0.39 & - \\
        $W$ Error (kg) & 9.45 & 0.34 & - \\
        $\boldsymbol{D}$ Error (cm) & 3.63 & 2.33 & 2.30 \\
        $\boldsymbol{C}$ Error (cm) & 3.78 & 1.26 & - \\
        \bottomrule
    \end{tabular}
    \caption{Comparison of reconstructed body mesh errors on TotalCapture~\cite{trumble2017total} using mean shape versus predicted shape parameters.}
    \label{table:com_shape}
\end{table}
\pref{table:com_shape} compares reconstructed body mesh on the TotalCapture dataset using the mean and predicted shape, both with the same estimated poses.
With the predicted shape, positional and mesh errors are nearly as low as those obtained using the ground truth shape parameters.
Height and weight estimates are accurate, with errors within 1 cm and 1 kg, respectively.
However, the mean error in circumferences $\boldsymbol{C}$ of the chest, waist, hips, wrists, knees, and head remains large, as relevant measurements cannot be inferred solely from inter-sensor distances.
Despite this limitation, the predicted shape still reconstructs a more realistic body than the mean shape model.

\noindent\textbf{Pose Estimator}\quad
\begin{table}[t]
    \small
    \centering
    \begin{tabular}{ccccc}
        \toprule
        \multirow{2}{*}{Method} & SIP Error &  Pos Error & SIP Error & Pos Error \\
               & (deg) & (cm) & (deg) & (cm) \\
        \midrule
        \multicolumn{1}{c}{} & 
        \multicolumn{2}{c}{TotalCapture} &
        \multicolumn{2}{c}{DIP-IMU}  \\
        \midrule
        {\footnotesize{$\boldsymbol{J} \rightarrow \hat\pose$}} & 9.00 & 4.36 & 14.21 & 3.29 \\
        {\footnotesize$\boldsymbol{S} \rightarrow \boldsymbol{J}\rightarrow\hat\pose$} & 12.15  & 5.70 & 17.35 & 4.45 \\
        {\footnotesize$\boldsymbol{P}_{G} \rightarrow \hat\pose$} & 9.25 & 3.85 & 13.63 & 2.85 \\
        {\footnotesize$\boldsymbol{S} \rightarrow \boldsymbol{P}_{G}\rightarrow\hat\pose$} & 11.13 & 4.96 & 15.91 & 3.91 \\ 
        {\footnotesize$\boldsymbol{P}_{R} \rightarrow \hat\pose$} & 9.89 & 4.42 & 13.77 & 3.22 \\
        {\footnotesize$\boldsymbol{S} \rightarrow \boldsymbol{P}_{R}\rightarrow\hat\pose$} & 11.57 & 5.44 & 15.74 & 4.17 \\ 
        \midrule
        {\footnotesize\textbf{Ours: $\boldsymbol{S} \rightarrow \hat\pose$}}  & \textbf{10.76} & \textbf{4.46} & \textbf{14.21} & \textbf{3.38} \\
        \bottomrule
    \end{tabular}
    \caption{Ablation study on pose estimator architecture.}
    \label{table:com_architecture}
\end{table}
We conduct an ablation study by expanding our pose estimator with intermediate layers to estimate all joint positions, $\boldsymbol{J}$, sensor global positions, $\boldsymbol{P}_G$, and sensor relative positions, $\boldsymbol{P}_{R}$, from sensor measurements $\boldsymbol{S} = (\hat{\a}, \boldsymbol{R}, \hat{\boldsymbol{d}})$.
These intermediate estimates are then combined with $\boldsymbol{S}$ to compute the final poses $\hat\pose$.
This layered structure follows the framework commonly used in previous methods~\cite{yi2021transpose, armani2024ultra}.
As shown in \pref{table:com_architecture}, while regressing $\hat\pose$ from ideal $\boldsymbol{J}$, $\boldsymbol{P}_{G}$, or $\boldsymbol{P}_{R}$ improves accuracy, introducing an intermediate layer, that is, $\boldsymbol{S} \rightarrow \{\boldsymbol{J}, \boldsymbol{P}_G, \boldsymbol{P}_R\}$, increases errors.
This error propagates through the network, ultimately degrading pose estimation performance.
This suggests that integrating IMU data with distance constraints may support a simpler architecture, reducing complexity while still maintaining accuracy.

\noindent\textbf{State Estimator}\quad
\begin{figure*}[t]
    \centering
    \includegraphics[width=.75\textwidth]{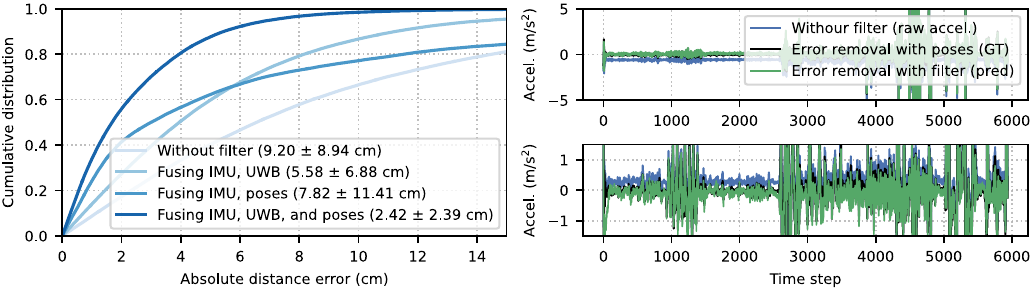}
    \caption{Cumulative distribution of distance error (left) and acceleration error reduction over time (right) for various fusion settings.}
    \label{fig:filter}
\end{figure*}
\begin{figure}[t]
    \centering
    \includegraphics[width=0.4\textwidth]{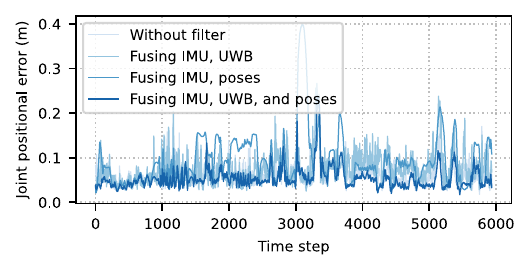}
    \caption{Joint positional error for different fusion settings.}
    \label{fig:pos_err}
\end{figure}
The state estimator is designed to: 1) mitigate errors in inter-sensor distances, 2) filter and stabilize accelerations, and 3) improve the accuracy of estimated poses.
To examine it we conduct an ablation study by varying fusion measurements on the TotalCapture~\cite{trumble2017total} with synthesized LOS-related noisy distances.
\pref{fig:filter} (left) shows cumulative distributions of absolute distance error for different fusion settings.
The raw distances show an average error of 9.20 cm with a standard deviation of 8.94 cm.
By fusing IMU, UWB, and pose data, this error is reduced to 2.42 cm with a standard deviation of 2.39 cm, demonstrating the effectiveness of data fusion for mitigating distance errors.
\pref{fig:filter} (right) shows the reduction in acceleration errors.
The state estimator converges within a few seconds and adapts gradually, aligning with IMU characteristics.
\pref{fig:pos_err} presents joint positional errors over time for different fusion setups.
Without filtering, positional error fluctuates with distance measurement quality.
Fusing IMU with either UWB or pose data alone offers limited improvement because distance and pose outliers remain.
The combined fusion of IMU, UWB, and pose data achieves the lowest positional error, validating that our state estimator refines pose accuracy by utilizing all available measurements.

\subsection{Qualitative Evaluation}
\begin{figure}
    \centering
    \includegraphics[width=.75\linewidth]{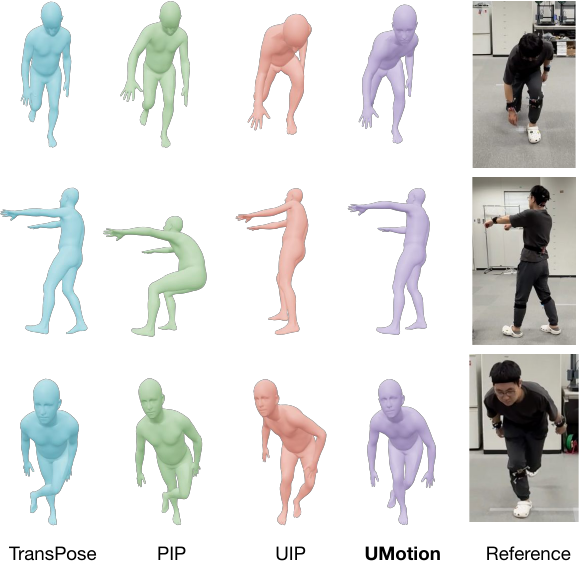}
    \caption{Qualitative comparison of pose estimates from the test
data collected using our developed prototype.}
    \label{fig:demo}
\end{figure}
We developed a prototype (see supplementary material for details) that integrates BNO086 IMU and DW3000 UWB sensors to demonstrate our proposed method.
\pref{fig:demo} presents visualizations comparing pose estimates for challenging motions, illustrating the improvements of our method in disambiguating poses over IMU-only methods and enhanced pose stability compared to UIP.

\section{Conclusion and Limitations}
In this work, we present UMotion, an uncertainty-driven framework for 3D human shape and pose estimation that integrates six IMU and UWB sensors within a state estimation system.
UMotion incorporates uncertainties from IMU, UWB, and pose estimates under individual body constraints to iteratively enhance confidence in both sensor measurements and pose accuracy.
Our experiments on synthetic and real-world datasets demonstrate that UMotion outperforms existing SOTA methods in pose accuracy, and effectively stabilizes sensor measurements and pose estimation.

However, our shape estimator requires adaptation to specific sensor configurations and conditions, and, with limited training data, it may struggle with unique body variations.
The state estimator also requires careful parameter tuning to reflect sensor characteristics accurately. 
Additionally, our pose estimator simplifies constraints, which may limit its performance. As future work, adding a physics-aware module could potentially enhance our method's robustness.

\section{Acknowledgement}
This work was supported by the Japan Science and Technology Agency under the Broadening Opportunities for Outstanding Young Researchers and Doctoral Students in Strategic Areas (BOOST) JPMJBS2423.

{
    \small
    \bibliographystyle{ieeenat_fullname}
    \bibliography{main}
}

\newpage
\setcounter{page}{1}
\setcounter{section}{0}
\maketitlesupplementary
\renewcommand{\thesection}{\Alph{section}}
\section{IMU-UWB Prototype}
We developed a prototype integrating the off-the-shelf CEVA BNO086 9-axis IMU and Qorvo DW3000 UWB sensors on a customized board.
An ESP32 microcontroller handles on-board data preprocessing and wireless transmission.
The BNO086 operates at 100~Hz, using an on-board sensor fusion algorithm to output linear acceleration (gravity-removed) in the sensor’s local coordinate frame $F^S$ and orientation relative to the initial frame.
The DW3000 sensors measure 15 inter-sensor distances at an average rate of 80~Hz, with a customized asymmetric double-sided two-way ranging protocol.
A time synchronization step is applied, followed by downsampling to align all measurements to 60 Hz.

\section{From IMU Readings to Input Measurements}
We follow the calibration procedures described in DIP~\cite{huang2018deep} and TransPose~\cite{yi2021transpose}, adapting them to suit the specific characteristics of the sensors used in our system.
\begin{figure}
    \centering
    \includegraphics[width=\linewidth]{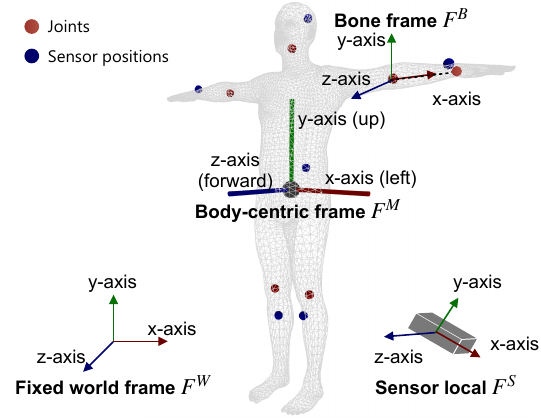}
    \caption{Overview of coordinate frames.}
    \label{fig:frame}
\end{figure}
\paragraph{Frame Definition}
IMU reading coordinate frame transformation is essential for aligning IMU data with the model input requirements.
As shown in~\pref{fig:frame}, the system operates with four types of coordinate frames:
\begin{itemize}
    \item Sensor local coordinate frame $F^S$: Each sensor has its own local frame, resulting in six frames in total.
    \item Fixed world frame $F^W$: For the BNO086, the fixed world frame corresponds to the first sensor frame upon power-up.
    Each sensor thus has its own $F^W$, totaling six frames.
    \item SMPL Body-centric frame $F^M$: A single frame per person, defined as Left-Up-Forward in this work.
    Motions are described relative to this fixed frame, which is initialized in the T-pose at the start of the motion sequence.
    \item Respective bone coordinate frame $F^B$: Each bone with a mounted IMU has its own coordinate frame, giving six frames in total.
\end{itemize}
In total, the system consists of 19 coordinate frames: one body-centric frame, $F^M$, and six groups of three frames each, comprising $F^{S,i}$, $F^{W,i}$, and $F^{B,i}$, where $i \in \{1, 2, \dots, 6\}$.

\paragraph{Problem Statement}
The IMU measures linear acceleration $\a^S$ in the sensor local frame $F^S$ and orientation $\boldsymbol{R}^{WS}$, which represents the rotation matrix that transforms vectors from the sensor frame $F^S$ to the fixed world frame $F^W$. When applied to a acceleration in $F^{S}$, $\a^{W} = \boldsymbol{R}^{WS}\a^S$ describes the acceleration's representation in $F^{W}$.
The inputs to the network are bone orientations relative to the body-centric frame, $\boldsymbol{R}^{MB}$, and linear accelerations in the body-centric frame, $\a^{M}$.
$\boldsymbol{R}^{MB}$ describes the rotation of each bone around the axes of the body-centric frame. These orientations also represent the global poses of the adjacent joints.
To align the IMU readings with the model input, we need to transform the sensor-local accelerations $\a^S$ into the body-centric frame $\a^M$, and the sensor-to-world orientation $\boldsymbol{R}^{WS}$ into the bone-to-body orientation $\boldsymbol{R}^{MB}$. These transformations are expressed as:
\begin{align}
    \boldsymbol{R}^{MB} &= \boldsymbol{R}^{MW}\boldsymbol{R}^{WS}\boldsymbol{R}^{SB}, \\
    \a^M &= \boldsymbol{R}^{MS}\boldsymbol{a}^S \nonumber\\
    &= \boldsymbol{R}^{MW}\boldsymbol{R}^{WS}\a^S.
\end{align}
The calibration process aims to determine $\boldsymbol{R}^{MW}$ and $\boldsymbol{R}^{SB}$ to enable these transformations.
\paragraph{Calculation of $\boldsymbol{R}^{MW}$} 
As shown in~\pref{fig:frame}, the body-centric frame $F^M$ is established as the Left-Up-Forward orientation of the initial T-pose at the start of the motion.
The fixed world frame of the BNO086 is defined as the first sensor frame after power-up.
To ensure consistency, we position all IMUs in the same initial orientation, aligning their initial sensor frames such that $F^{S, 1}_{\text{init}} = \dots =  F^{S, 6}_{\text{init}} = F^{W, 1} = F^{W,2} = \dots = F^{W,6}$.
To simplify computation, we align the axes of the $F^{S}_{\text{init}}$ and $F^{W}$ with the corresponding axes of $F^M$ or use a known transformation.
For example, we position the IMU with its x-axis pointing left, y-axis pointing up, and z-axis pointing forward in the real world.
This alignment defines $\boldsymbol{R}^{MW}$.
In cases where $F^W$ is aligned with $F^M$, $\boldsymbol{R}^{MW} = \mathbf{I}$.

\paragraph{Calculation of $\boldsymbol{R}^{SB}$}
Next, we mount IMUs onto the corresponding body part in arbitrary orientations.
The subject is then instructed to remain still in a T-pose for several seconds.
In this pose, the orientation of bone frame relative to the SMPL body-centric frame is zero, meaning $\boldsymbol{R}^{MB}_{\text{T-pose}} = \mathbf{I}$.
Thus, given the measured average orientation of the IMU in T-pose, $\bar{\boldsymbol{R}}^{WS}_{\text{T-pose}}$, we have
\begin{align}
\boldsymbol{R}^{MB}_{\text{T-pose}} &= \boldsymbol{R}^{MW}\bar{\boldsymbol{R}}^{WS}_{\text{T-pose}}\boldsymbol{R}^{SB},\\
\boldsymbol{R}^{SB} &= \text{inv}(\boldsymbol{R}^{MW}\bar{\boldsymbol{R}}^{WS}_{\text{T-pose}})\boldsymbol{R}^{MB}_{\text{T-pose}}, \\
\boldsymbol{R}^{SB} &= \text{inv}(\bar{\boldsymbol{R}}^{WS}_{\text{T-pose}}).
\end{align}

\section{Ranging Protocol}
\begin{figure}
    \centering
    \includegraphics[width=\linewidth]{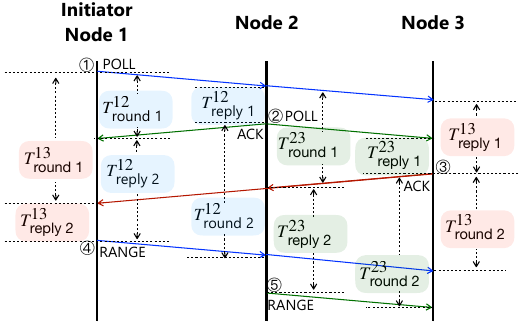}
    \caption{Ranging transaction with three devices. Timestamps to resolve time-of-fight are included in the UWB message payload and thus broadcast to all network participants.}
    \label{fig:twr}
\end{figure}
We implemented an efficient distance matrix ranging method based on asymmetric double-sided two-way ranging (ADS-TWR) protocol~\cite{mclaughlin2019asymmetric}.
Compared to the standard two-way ranging protocol, ADS-TWR minimizes the impact of clock drift and synchronization errors.
\pref{fig:twr} illustrates an example with three sensors.
One sensor is designated as the initiator and transmits a POLL signal.
Subsequently, other sensors sequentially act as transmitters, sending POLL signals to the remaining sensors after receiving POLL signals from all preceding sensors in order.
These POLL signals simultaneously serve as ACK signals for the previous sensors, streamlining communication.
This efficient broadcasting strategy reduces the number of transmitted signals from 45 (calculated as 15 pairs, each requiring 3 transmissions) to just 11.
A sequence of timestamps is recorded during this process to measure the time-of-flight (ToF), $T$, between sensor pairs. $T$ is determined using the formula:
\begin{equation}
T = \frac{T_{\text{round 1}} \times T_{\text{round 2}} - T_{\text{reply 1}} \times T_{\text{reply 2}}}{T_{\text{round 1}} + T_{\text{round 2}} +T_{\text{reply 1}} +T_{\text{reply 2}}}.
\end{equation}
The corresponding distance, $d$, between the sensor pairs is then calculated as:
\begin{equation}
    d = cT,
\end{equation}
where $c$ represents the speed of light in vacuum.

\section{Line of Sight Simulation}
\begin{figure}
    \centering
    \includegraphics[width=\linewidth]{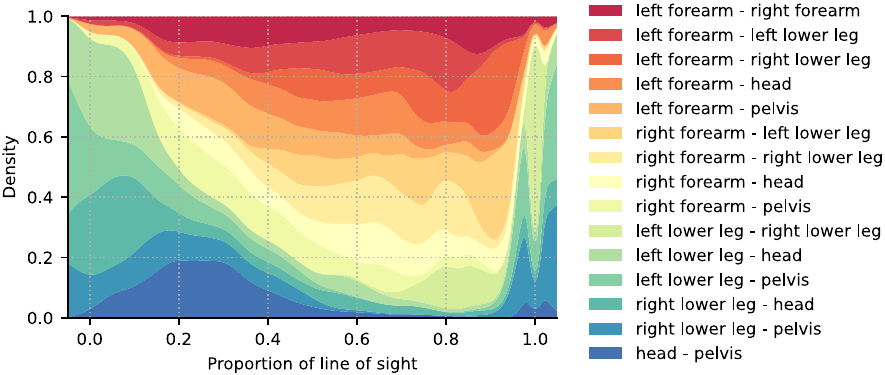}
    \caption{Stacked density plot showing the proportion relative to the total distribution of LOS availability for inter-sensor distances across different sensor pairs. }
    \label{fig:los}
\end{figure}
One challenge in using body-worn UWB sensors for tracking inter-sensor distances is body occlusion, which degrades measurement accuracy~\cite{armani2024ultra}.
To address this, we simulate line-of-sight (LOS) conditions to learn the distribution of the occlusion on TotalCapture dataset~\cite{trumble2017total}.
The simulation utilizes the SMPL body model~\cite{loper2023smpl} to calculate LOS and non-line-of-sight (NLOS) conditions based on different poses.
The visibility of each sensor pair is determined by tracing straight-line paths between them and checking for intersections with the body mesh.
We employ the Möller–Trumbore intersection algorithm to identify these intersections.
The LOS proportion is then calculated as the total length of unobstructed (LOS) segments divided by the entire distance.

\pref{fig:los} shows a stacked density plot of LOS proportions across 15 sensor pairs, representing the relative contribution of each sensor pair to the total distribution of LOS proportions.
For a given LOS proportion, the stacked regions indicate how frequently different sensor pairs contribute to that proportion.
It reveals that pairs such as ``lower leg - pelvis'' and ``lower leg - head'' exhibit consistently low LOS availability due to frequent occlusion caused by body movement and overlapping limbs.
Accordingly, the corresponding distance measurements are unreliable and could not be effectively used for pose estimation or measurement filtering.
This analysis highlights the varying reliability of UWB measurements across sensor pairs, offering guidelines for weighting measurement uncertainties in our state estimation framework.
\begin{figure}
    \centering
    \includegraphics[width=\linewidth]{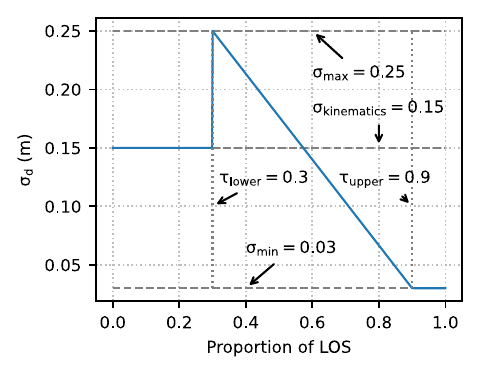}
    \caption{Example of the distance error model based on the LOS proportions for the sensors used in our system.}
    \label{fig:error_model}
\end{figure}

\noindent\textbf{Distance Error Model}\quad In this work, we simplify the standard deviation of distance measurements, $\sigma_d$, as a function of the LOS proportion, $l$, as follows:
\begin{equation}
    \sigma_d = 
    \begin{cases} 
\sigma_{\text{min}}, & \text{if } l \geq \tau_{\text{upper}}, \\
\sigma_{\text{kinematics}}, & \text{if } l < \tau_{\text{lower}}, \\
(\sigma_{\text{max}} - \sigma_{\text{min}})\frac{(\tau_{\text{upper}} - l)}{\tau_{\text{upper}} - \tau_{\text{lower}}}  + \sigma_{\text{min}}, & \text{otherwise},
\end{cases}
\end{equation}
where $\tau_{\text{upper}}$ and $\tau_{\text{lower}}$ are LOS proportion thresholds, and $\sigma_{\text{min}}$ and $\sigma_{\text{max}}$ represent the minimum and maximum noise parameters for the distance standard deviation. When the LOS proportion falls below $\tau_{\text{lower}}$, the distance measurement is replaced with one derived from kinematics, with an associated standard deviation of $\sigma_{\text{kinematics}}$.
\pref{fig:error_model} provides an example of this model based on our selected sensors.
The parameters may vary depending on the specific sensors used.

\section{Discussions on Predicted Uncertainty}
\begin{figure}
    \centering
    \includegraphics[width=0.9\linewidth]{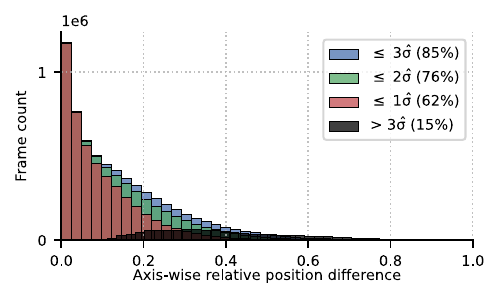}
    \caption{Histogram of axis-wise relative position differences, illustrating the alignment between predicted uncertainty and true errors.}
    \label{fig:uncertainty}
\end{figure}
To assess the correctness of the predicted uncertainty, we analyze the transformed axis-wise relative position error distributions.
We calculate distance errors given predicted poses and compare them with the distance uncertainty into which the predicted pose uncertainty is converted.
\pref{fig:uncertainty} shows the proportion of frame counts within different confidence intervals.
The results indicate that the predicted uncertainty aligns well with actual errors for smaller deviations, with 85\% of predictions falling within $3\sigma$.
However, for larger errors, the predicted uncertainty tends to be underestimated.

\section{Implementation Details}
We train the pose estimator using synthesized data from the AMASS dataset without integrating the state estimator.
We apply noise only to the synthesized distances, while synthesized IMU data remains noise-free.
In the state estimator, the process noise covariance $\mathbf{Q}$ is determined using Allan variance analysis with a noise propagation model.
The observation noise covariance $\mathbf{R}_1$ follows our distance error model, while $\mathbf{R}_3$ is derived from predicted poses via the unscented transformation.
To mitigate overconfidence in high-error scenarios, we scale $\mathbf{R}_3$ by a factor of 10 for improved stability.

\section{Ablation on Shape Estimator}
\begin{table}[b]
    \small
    \centering
    \begin{tabular}{cccccc}
        \toprule
        & \multicolumn{5}{c}{Mean absolute error}\\
        & Mesh (mm) & $H$ (mm) & $W$ (kg) & $\boldsymbol{D}$ (mm) & $\boldsymbol{C}$ (mm) \\
        \midrule
        H & 12.10  & 1.11 & 3.77 & 10.62 & 21.08 \\
        W & 23.45 & 58.70 & 0.28 & 31.69 & 16.11\\
        D & 6.14 & 2.67 & 4.47 & 0.9 & 22.62 \\
        HW & 10.40 & 1.2 & 0.19 & 11.30 & 13.34 \\
        HD & 6.30 & 1.83 & 4.10 & 1.34 & 21.08 \\
        WD & 4.31 & 3.37 & 1.03 & 1.14 & 13.26 \\
        HWD & 4.72 & 3.89 & 0.35 & 2.09 & 12.76 \\
        \bottomrule
    \end{tabular}
    \caption{Comparison of reconstructed T-pose mesh errors on TotalCapture~\cite{trumble2017total} using different sources of anthropometric data.}
    \label{table:ablation_shape}
\end{table}
To evaluate the impact of different anthropometric data on shape estimation, we conduct an ablation study using the TotalCapture dataset.
Table~\ref{table:ablation_shape} presents the mean absolute error of the reconstructed T-pose mesh under different subsets of anthropometric inputs.
Since circumferences are not directly observed, their errors remain the highest across all conditions.
Using only height (H) or weight (W) results in relatively large distance and mesh errors, demonstrating that these individual measurements alone do not sufficiently constrain body shape.
Combining height and weight (HW) improves shape estimation, leading to slight reductions in mesh errors.
Incorporating inter-sensor distances (D) provides better constraints on body proportions, further reducing mesh and distance errors.

\section{Ablation on State Estimator}
\begin{figure*}[t]
    \centering
    \includegraphics[width=0.9\textwidth]{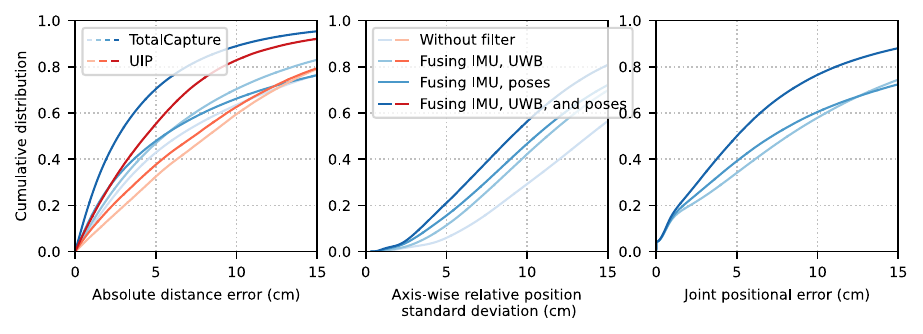}
    \caption{Cumulative distribution of distance error (left), predicted relative position standard deviation (middle), and joint positional error (right) for various fusion settings.}
    \label{fig:state_estimator_ablation}
\end{figure*}
We compare absolute distance error, predicted uncertainty, and joint positional error across various configurations on TotalCapture and UIP datasets to evaluate the impact of different fusion strategies.
\pref{fig:state_estimator_ablation} (left) illustrates the cumulative distribution of absolute distance errors.
Incorporating IMU and UWB fusion reduces distance errors, and the addition of pose information further improves accuracy.
This demonstrates that integrating multiple sensing modalities enhances distance estimation by leveraging complementary information.
\pref{fig:state_estimator_ablation} (middle) shows the axis-wise relative position standard deviations, evaluating the effect of different information on the predicted uncertainty.
The results indicate that the full fusion model, i.e., IMU, UWB, and poses, improves the consistency of uncertainty estimation, resulting in the most confident predictions.
\pref{fig:state_estimator_ablation} (right) evaluates the cumulative distribution of joint positional errors.
Compared to the unfiltered case, fusing IMU and UWB data reduces error, while incorporating pose constraints further improves tracking performance.
These results demonstrate that jointly fusing IMU, UWB, and pose constraints improves distance accuracy, refines uncertainty estimation, and reduces joint positional errors.

\end{document}